\documentclass[11pt]{article}
\usepackage{mathrsfs}
\usepackage{amssymb,bm}
\usepackage{color}

\usepackage{hyperref}
\usepackage{amsthm,float}
\usepackage{amsmath}
\usepackage{algorithm}
\usepackage{algorithmic}

\usepackage[a4paper, total={6in, 8in}]{geometry}

\usepackage{graphicx,tikz}
\usepackage{subfigure}

 \theoremstyle{definition}

\let\<=\langle
\let\>=\rangle
\def\E{{\mathbb E}}
\def\~#1{{\text{\sf#1}}}

\let\-=\mathbf
\let\^=\hat
\def\@#1{{\cal #1}}

\def\COV{\mathrm{COV}}

\def\u0{u_{0}}
\def\l2{l_{2}}

\begin{document}
\title{Gaussian Process Regression for Uncertainty Quantification: An Introductory Tutorial}
\author{Jinglai Li\footnote{School of Mathematics,
University of Birmingham, Birmingham, B15 2TT, United Kingdom. Email: j.li.10@bham.ac.uk.}\quad and\quad 
Hongqiao Wang\footnote{School of Mathematics and Statistics, Central South University, Changsha, 410083, China. 
Email: hongqiao.wang@csu.edu.cn}}

\date{}

\maketitle

\begin{abstract}

Uncertainty Quantification (UQ) is essential for the reliable application of computational models in engineering and science. Among surrogate modeling techniques, Gaussian Process Regression (GPR) is particularly valuable for its non-parametric flexibility and inherent probabilistic output.
This paper presents an introductory review of GPR-based methodologies within the context of UQ. 
We begin with an introduction to UQ and outline its key tasks, including uncertainty propagation, risk estimation, optimization under uncertainty, parameter estimation, and sensitivity analysis.
We then introduce Gaussian Processes as a surrogate modeling technique, detailing their formulation, choice of covariance kernels, hyperparameter estimation, and active learning strategies for efficient data acquisition. The tutorial further explores how GPR can be applied to different UQ tasks, including Bayesian quadrature for uncertainty propagation, active learning-based risk estimation, Bayesian optimization for optimization under uncertainty, and surrogate-based sensitivity analysis.
Throughout, we emphasize how to leverage the unique formulation of GP for these UQ tasks, rather than simply using it as 
a standard surrogate model.
This work offers a comprehensive guide and unified framework for researchers seeking to rigorously apply probabilistic modeling to complex computational systems.

\end{abstract}

\section{Introduction and background}
We consider the applications of the Gaussian Process regression (GPR) in the field of uncertainty quantification (UQ). 
This work is not meant to provide an exhaustive review of the subject; instead, our aim is to cover the foundational concepts, with a particular focus on beginners.

\subsection{What is Uncertainty Quantification}
Computer models and simulations play an essential role in many real-world applications, ranging from engineering design,
scientific exploration, to public policy making. 
Regardless of how accurate the underlying computer models are, uncertainty in their simulation results is inevitable. 
The primary source of uncertainty in computer models arises from factors such as inherent 
randomness in the modeled system, limited accuracy of input data, and simplifications made in representing complex real-world processes.
On the other hand, as many important decisions are made based on the simulation results of the computer models, 
a critical question to ask is
how credible the simulation results are. 
Or oppositely, \textit{how much uncertainty in the results?}
Uncertainty quantification (UQ) is a science that intends to answer these questions. 
More precisely UQ deals with assessing, characterising and managing uncertainty in computer models and simulations. 
It is especially important in situations where accurate predictions are crucial but difficult to obtain. 

There are two primary types of uncertainty:
\begin{itemize}
\item Aleatory Uncertainty: Inherent randomness in the system of interest, often due to natural variations that can not be characterized with certainty.
For instance, many real-world engineering or physical systems are perturbed by  external forces that are inherently random. 
Aleatory uncertainty cannot be eliminated or reduced, regardless of the amount of information or data available.
\item Epistemic Uncertainty: Arises from lack of knowledge, incomplete data, or model simplifications.
A good example of such type of uncertainty is the unspecified parameters in a computer model. 
Epistemic uncertainty can be reduced by gaining more and better information about the system, e.g. collecting more data  or 
improving model accuracy. 
\end{itemize}
In most practical problems, both aleatory and epistemic uncertainty
are present and need to be dealt with. 
Also the distinction between epistemic and aleatoric uncertainty is quite ambiguous and even subjective,
and as such  we do not seek to distinguish these two types of uncertainty sources in UQ problems.
This is an oversimplified introduction to UQ and we refer to \cite{ghanem2017handbook,smith2024uncertainty,sullivan2015introduction} for more a comprehensive discussion.

\subsection{UQ tasks}
Depending on different application purposes, there are a variety of UQ tasks.
In what follows we introduce some representative problems or tasks related to UQ.

\subsubsection{Uncertainty propagation (UP)}  The goal of this task is to quantify how uncertainties in input parameters or variables of a system propagate through mathematical models  to affect the  output or predicted results.
Let $y=f(\-x)$ represent the underlying computer model, where
$x$ is the random input variables and $y$ is the  system output. 
Both $\-x$ and $y$ can be multivariate. 
The input $\-x$ is typically in the class of aleatory uncertainty, representing inherent uncertainty in the system, but it can be epistemic too. 
One is interested in certain statistical information of the model output $y$, and most commonly 
its mean and variance. Monte Carlo (MC) simulations are often used to estimate the statistical quantities. For example, the mean of $y$ can be estimated as,
\begin{equation}
\mathbb{E}[y] = \int f(\-x) \pi(\-x)dx \approx \frac1N \sum_{n=1}^Nf(\-x_n), \label{e:upmc}
\end{equation} 
where $\pi(\cdot)$ is the distribution density of $\-x$ and $\-x_1,...,\-x_N$ are i.i.d. samples drawn from it. 
Its (co)-variance or other statistical quantities such as higher-order moments can be estimated similarly. 
As one can see, evaluating these statistical quantities can also be formulated as numerical integration problems, 
and therefore, methods for numerical integration, such as the sparse grid techniques or quasi-MC can be used as well. 
Nevertheless, a large number of forward model (i.e., $f(\cdot)$) evaluations are needed in all these methods.

\subsubsection{Risk estimation (RE)} \label{sec:re}
In many engineering systems,  some random events may cause serious consequences,  with the most severe being system failures. 
The occurrence of these events is regarded as the risk associated with the system, and the goal of risk assessment is to 
assess how likely such a random event may occur. 
The formulation of RE problems 
is similar to uncertainty propagation, except that in this type of problems, 
one is interested in the probability of a certain random event.
More specifically the event of interest is usually defined via the outputs
of the underlying model:
\[P_A=\mathbb{P}[f(\-x)\in A],\]
where $A$ represents the random event associated with risk.
In many engineering applications, the risk is typically interpreted 
as the occurrence of certain system failures and in this case the problem is also called failure probability estimation. 
For convenience's sake, we often formulate the problem in a way that the output of $f(\-x)$ 
is a scalar\footnote{Note that in this case, the function $f(x)$ is 
defined differently from that used in UP.}.
the event of risk is often defined as $f(\-x)<0$, which implies that
$A = (-\infty, 0)$, 
and  in this case the function $f(\-x)$ is often referred to as the limit state function. 
Realistically, the event $A$ is usually rare, 
which means the aforementioned probability is extremely small.
As a result evaluating $P_A$ is particularly challenging and 
advanced sampling schemes are often needed. 
We here present Importance Sampling (IS) as an example for such sampling schemes. 
Similar to UP, the probability of interest can be estimated via MC simulation:
\begin{equation}
    P_A = \mathbb{E}_\pi[I_A(f(\-x_n))]\approx \hat{P}_{MC} =  \frac1N \sum_{n=1}^N I_A(f(\-x_n)),
\end{equation}
where $$I_A (y) = \left\{
\begin{matrix}1,\quad y\in A;\\0,\quad y\notin A;
\end{matrix} \right.$$
and $\-x_1,...,\-x_N$ are independent and identically distributed (i.i.d.)~samples drawn from $\pi(\-x)$.
A difficulty with MC is that, when the sought probability is rare, only a very small fraction of samples 
result in the event of interest, i.e., $f(\-x)\in A$. 
The technique of IS \cite{rubinstein2016simulation} can be used to address the issue. 
Namely, given an alternative distribution $\pi^*(x)$, IS writes probability $P_A$ as 
\begin{equation}
    P_A = \mathbb{E}_{\pi^*} [ I_A(f(\-x))w(\-x)] \quad \mathrm{with}\quad w(\-x) = \pi(\-x)/\pi^*(\-x), 
\end{equation}
which can be estimated via 
\begin{equation} 
\hat{P}_{IS} = \frac1N\sum_{n=1}^N I_A(f(\-x_n))w(\-x_n),
\end{equation}
where now the samples $\-x_1,...,\-x_N$ are drawn from the alternative distribution $\pi^*(x)$. \
IS is a variance reduction technique, and its performance critically depends on the choice of $\pi^*$.
A large amount of works have been done on designing effective IS distribution $\pi^*(\cdot)$ (see e.g.,  \cite{rubinstein2016simulation}), which are not discussed here.

\subsubsection{Optimization under Uncertainty (OU)} \label{sec:ou}

Optimization under uncertainty deals with finding the best decision in the presence of uncertain parameters. 
It is often modeled using stochastic optimization techniques.
Consider the following optimization problem under uncertainty:
\begin{subequations}\label{e:ouu}
\begin{align}
\text{Minimize} \quad & E_{\xi}[f(x, \xi)] \\
\text{subject to} \quad & E_{\xi}[g_i(x, \xi)] \leq 0, \quad i = 1, \ldots, n_{g} \\
&E_{\xi}[ h_j(x, \xi)] = 0, \quad j = 1, \ldots, n_h
\end{align}
\end{subequations}
where:
\begin{itemize}
\item $x$ is the decision variable.
\item $\xi$ represents uncertain factors, modeled as random variables following distribution $\pi_{\xi}(\cdot)$.
\item $f(x, \xi)$ is the objective function.
\item $g_i(x, \xi)$ and $h_j(x, \xi)$ are inequality and equality constraints, respectively.
\end{itemize}

There are a large number of stochastic optimization algorithms to solve problem~\eqref{e:ouu}, and here we use the
common sample average approximation (SAA) method \cite{kim2015guide} as an example. 
The basic idea of SAA is to approximates the expected values of the objective and the constraint functions by averaging a finite number of sample scenarios,
resulting in the so-called stochastic counterpart of Eq.~\eqref{e:ouu}: 
\begin{subequations}\label{e:sc}
\begin{align}
\text{Minimize}\quad  \hat{f}(x)&= \frac1N\sum_{n=1}^Nf(x, \xi_n) \\
\text{subject to}\quad  \hat{g}(x)&= \frac1N\sum_{n=1}^N g_i(x, \xi_n) \leq 0, \quad i = 1, \ldots, n_g \\
\hat{h}(x)&=\frac1N\sum_{n=1}^N h_j(x, \xi_n) = 0,  j = 1, \ldots, n_h, 
\end{align}
\end{subequations}
where $\xi_1,...,\xi_N$ are i.i.d. samples drawn from distribution $\pi_{\xi}$. 
 Problem~\eqref{e:sc} is a standard constrained optimization problem,
and therefore can be solved with the usual optimization techniques. 
It should be noted that in reality the objective function and/or the constraints
typically depend on the underlying mathematical model.

\subsubsection{Parameter estimation (PE)} \label{sec:pe}
There are some parameters in the computer models that are unspecified
and need to be inferred from data. 
Due to observation noise and the model errors, 
it is usually impossible to  estimate the parameters with complete accuracy,
and the estimation uncertainty needs to be characterised. 
To this end, the Bayesian method \cite{gelman2013bayesian} is often used to estimate the parameters,
thanks to its ability to characterise the uncertainty in the estimation results. 
Mathematically, we assume that the parameter of interest $x$ is related to
the observation data $y$ via a likelihood function $\pi(\-y|\-x)$. 
A prior distribution $\pi(\-x)$ is imposed on the parameter of interest $x$ which represents the prior knowledge or information on $x$.
The posterior distribution is then calculated vie the Bayes' formula:
\begin{equation}
    \pi(\-x|\-y)=\frac{\pi(y|\-x)\pi(\-x)}{\pi(\-y)},\label{e:poster}
\end{equation}
where $\pi(\-y)$ can be understood as the normalization constant. 
In reality the underlying mathematical model is usually incorporated in the likelihood function. 
For example, in the common inverse problem setup, the data $\-y$ is linked to the parameters  $\-x$ via
\[\-y= f(\-x)+\eta,\]
where $f(x)$ represents the underlying mathematical model 
and $\eta$ is the observation noise
following distribution $\pi_{\eta}(\cdot)$. 
In this case, the likelihood is given by
\[\pi(\-y|\-x) = \pi_{\eta}(\-y-f(\-x)).\]
In practice, it is usually not possible to compute the posterior distribution~\eqref{e:poster} directly.
Instead one often draw samples from it via a sampling scheme. 
In this regard, the Markov Chain Monte Carlo (MCMC) methods \cite{robert2004monte} are commonly used, and the very basic Metropolis-Hastings
MCMC performs the 
following iterations (assuming at the $n$-th step): 
\begin{enumerate}
\item draw $\-x^*\sim q(\cdot|\-x_n)$;
\item let $a=\min\{1,\frac{\pi(\-x^*|\-y)}{\pi(\-x_n|\-y)}\frac{q(\-x_n|\-x^*)}{q(x^*|\-x_n)}\}$;
\item draw $u\sim U[0,1]$;
\item if $u<a$ let $\-x_{n+1}=\-x^*$; otherwise let $\-x_{n+1}=\-x_n$;
\end{enumerate}
where $\-x_n$ is the current state and $q(\cdot|\cdot)$ is a pre-selected proposal distribution. 
Typically the MCMC methods require a rather large number of iterations to converge to the posterior distribution, and as 
one can see, either iteration involves the evaluation of the forward model $f(\cdot)$. 

In certain context, the task of model calibration can be also related to parameter estimation. 
Namely, suppose we have a mathematical model with certain parameters that can be adjusted. 
On the other hand, we have observations of the system/model outputs. The parameters are tuned so that the model outputs agree with the observed data. 
One can see here that the model calibration problem can be formulated as parameter estimation, 
where the main difference lies on the purposes of the two problems: 
in parameter estimation, one is mainly interested in obtaining the true values of the parameters which may be useful for other purposes,
while in model calibration, the focus is that the model is correctly calibrated so that it can produce the correct outputs.

%

\subsubsection{Sensitivity Analysis (SA)} \label{sec:sa}

Sensitivity analysis (SA) \cite{saltelli2008global} is a task to study the relationship between input variables and output responses of a mathematical model. It aims to quantify the relative importance of input factors in contributing to the variability of the model outputs. SA is particularly useful in understanding the behavior of complex systems and in identifying parameters that are critical for system performance.
Specifically we consider the mathematical model $y= f(\-x)$ where $x$ represents the model parameters,  and $y$ is a scalar characterizing the system performance. 
Assume that $\-x$ is a $d$-dimensional random variable $\-x=[x_1,...,x_d]$ and we want to understand which components of $\-x$
are the most influential on the system performance characterised by $y$. 
In other word, we want to know how sensitive the output $y$ is to each $x_i$ for $i=1,...,d$. 
SA methods can be classified as local SA and global SA methods. 
Simply put, local SA aims to extract the sensitivity information at a particular location of the input space,
while the global SA examines the entire space of the input variables rather than a specific point. 
We here mainly focus on the global SA (GSA) methods \cite{saltelli2008global}. 

One commonly used approach for the GSA is the computation of sensitivity indices. These indices provide a measure of how much the output of a model changes in response to variations in input factors. The first-order sensitivity index of $x_i$ is defined as:
\begin{equation}
S_i = \frac{\mathrm{Var}(\E(y|x_i))}{\mathrm{Var}(y)}
\end{equation}
where $\E(y|x_i)$ is the conditional expectation of $y$ given $x_i$, and $\mathrm{Var}(\cdot)$ denotes the variance. The first-order sensitivity index measures the proportion of the total output variance that can be attributed to variations in $x_i$ alone.
Similarly, the total sensitivity index of $x_i$ is given by:
\begin{equation}
T_i = \frac{\E(\mathrm{Var}(y|x_i))}{\mathrm{Var}(y)}
\end{equation}
which quantifies the total effect of $x_i$ on the output, including both direct and indirect effects through interactions with other input variables.
These sensitivity indices are also called Sobol' indices.

To estimate sensitivity indices $S_i$ and $T_i$, Monte Carlo sampling is often employed. 
Here for simplicity we assume that  the random variables are independent in SA problems, and therefore the probability distribution of 
\( \mathbf{x} \) can be written as, 
 \[ \pi(\mathbf{x})= \prod_{i=1}^d \pi_i(x_i). \]
 We begin by generating two sets of \( n \) independent samples from \( \pi(\mathbf{x}) \), . 
These two sets of samples are organized into two \( n \times d \) matrices, \( A \) and \( B \), where each row represents an individual sample from the respective sets.
 Then, construct a mixed matrix \( A_B^{(i)} \) by replacing the \( i \)-th column of \( A \) with the corresponding column from \( B \).
The model outputs are then evaluated for all matrices:
\begin{equation}
Y_A = f(A), \quad Y_B = f(B), \quad Y_{A_B^{(i)}} = f(A_B^{(i)}). \label{e:eval_f}
\end{equation}
The total variance \( \mathrm{Var}(y)\) is estimated as:
\begin{subequations} \label{eq:mcsa}
\begin{equation}
\hat{V}_y = \frac{1}{n} \sum_{j=1}^n Y_A[j]^2 - \left(\frac{1}{n} \sum_{j=1}^n Y_A[j]\right)^2.
\end{equation}
The first-order variance $\mathrm{Var}(\E(y|x_i))$ is estimated as:
\begin{equation}
\hat{V}_i = \frac{1}{n} \sum_{j=1}^n Y_A[j] \cdot Y_{A_B^{(i)}}[j] - \left(\frac{1}{n} \sum_{j=1}^n Y_A[j]\right)^2.
\end{equation}
The first-order Sobol' index is  then computed as:
\begin{equation}
S_i = \frac{\hat{V}_i}{\hat{V}_y}.
\end{equation}
Similarly we can compute the total sensitivity index, which includes all interactions involving \( X_i \), is:
\begin{equation}
{T}_i = 1 - \frac{\hat{V}_{\sim i}}{\hat{V}_y},
\end{equation}
where 
\begin{equation}
\hat{V}_{\sim i} = \frac{1}{n} \sum_{j=1}^n Y_B[j] \cdot (Y_{A_B^{(i)}}[j] - Y_A[j]).
\end{equation}
\end{subequations}
While straightforward, the MC method requires a large number of evaluations of \( f(x) \).
Specifically, a total of $(d+2)n$ model evaluations are needed to compute the indices for all factors.

\subsection{A toy example}
\begin{figure}[h]
\begin{center}
\begin{tikzpicture}
    \draw[thick] (-3,0) -- (3,0);
    
    \draw[thick] (0,0) -- (2,-4);
    
    \filldraw[fill=gray] (2,-4) circle (0.3); 
    \node at (2,-4) {$m$};
    
    \node at (1.2,-2) [above] {$L$};

    \draw[dashed,->] (2,-4.1) -- (2,-5) node[midway,right] {$g$};

    \draw[dashed] (0,0) -- (0,-4);
    \draw[<->] (0,-1.5) arc[start angle=-90, end angle=-63.43, radius=1.5];
    \node at (0.4,-2) {$\theta$};
\end{tikzpicture}
\end{center}
\caption{A schematic illustration of the simple pendulum.} \label{f:pend}
\end{figure}
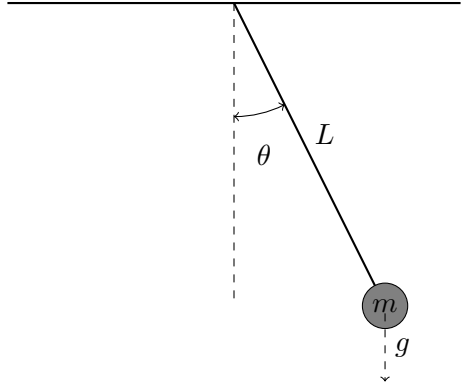

A simple pendulum, illustrated in Fig.~\ref{f:pend}, consists of a point mass (called the bob) attached to the end of a lightweight, inextensible string or rod of fixed length, which is allowed to swing freely under the influence of gravity. The motion of the simple pendulum can be described by the following model:
\begin{equation}
\frac{d^2\theta}{dt^2} + \frac{g}{L} \sin(\theta) = b(t),\quad\theta(0)=\theta_0, \label{e:pend}
\end{equation}
where \( \theta \) is the angle between the string and the vertical direction, $b(t)$ is a combination of damping and driving forces, 
\( L \) is the length of the rod, $g$ is the gravity acceleration and $t$ is time. 

The possible uncertainty factors in this model include:
\begin{itemize}
    \item \textbf{Model uncertainty}: The model is derived under some assumptions that are not exactly accurate, such as the bob being a point mass, and the rod being massless and inextensible.
    \item \textbf{Data uncertainty}: Length \( L \) needs to be measured, which is subject to measurement error.
    \item \textbf{Parameter uncertainty}: The exact value of $g$ is not known.
    \item \textbf{Inherent uncertainty}: Initial condition can not be controlled perfectly, and the forcing term $b(t)$ is also inherently random.  
     \item \textbf{Computational uncertainty}: The numerical implementation of the mathematical model suffers from numerical errors and approximations.
\end{itemize}

Below are some toy UQ problems formulated within the pendulum model described by Eq.~\eqref{e:pend},
where for simplicity the forcing term $b(t)$ is assumed to be zero in all the examples. 
\begin{itemize}
\item UP: We assume  that the  sources of uncertainty in the model are the initial condition $\theta_0$, the rod length $L$ and the gravity acceleration $g$,
and all the other model specifications are fixed.  
Moreover, we assume that the three random parameters are independent from each other and their distributions are 
$\pi_{\theta}(\cdot)$, $\pi_L(\cdot)$ and $\pi_g(\cdot)$ respectively. 
 Our goal is to determine the expectation and variance of the angle \( \theta(t) \) at a given time \( T \).

\item RE: The assumptions are the same as those in the UP example. 
The event of interest is defined as that $\theta(T)$ (once again for a fixed time $T$) is larger than a prescribed threshold value $\theta_{\max}$ and 
we want to estimate $\mathbb{P}[y\geq \theta_{\max}]$. 
\item PE: We assume that $\theta_0$, $L$ and $g$ are unknown parameters: they are independent and the prior distributions of them are 
$\pi_{\theta}(\cdot)$, $\pi_L(\cdot)$ and $\pi_g(\cdot)$ respectively. 
 Suppose we observe the angle \( \theta(t) \) at a number of discrete time points \( t_1, t_2, \ldots, t_m \), with observations \( y_i = \theta(t_i) + \epsilon_i \), where \( \epsilon_i \) represents measurement noise with distribution $\pi_{\epsilon}(\cdot)$.  
The goal is to estimate $\theta_0$, $L$ and \( g \) from these observations via Bayesian inference. 
\item SA: The assumptions are the same as those in the UP example, and the quantity of interest is 
$\theta(T)$ for a fixed time $T$. 
Through GSA (i.e., computing the sensitivity indices) we want to assess which parameter among $\theta_0$, $L$ and $g$ is 
most influential on the output $\theta(T)$. 
\item OU: First we make the same assumptions as those in the UP example,
and the quantify of interest is also $\theta(T)$ for a fixed $T$. 
Furthermore, we assume that  $\pi_{\theta}(\cdot)$ is a normal distribution with mean $\mu_{\theta}$ and variance $\sigma^2_\theta$.
We consider a control problem: loosely speaking, we want $\theta(T)$ to be close to a prescribed target value $\theta^*$ while its angular velocity is non-negative. When uncertainty is taken into account, the mathematical formulation of the problem is: 
\[ \min_{\mu_\theta} \E[(\theta(T)-\theta^*)^2]\quad \text{s.t.}\quad \E[\frac{d\theta}{dt}(T)]\geq 0,
\]
where the expectation is taken with respect to $\theta_0$, $L$ and $g$.

\end{itemize}

\subsection{Surrogate modeling}
Most of the aforementioned UQ tasks are computationally intensive
and the reason is two-fold. 
First, as has been discussed, these problems are often solved with the MC type of methods, which requires 
to perform a large number of simulations of the underlying mathematical models.
Second, for many real-world systems, the underlying mathematical models are highly complex and 
computationally intensive: for example, many of them are modelled by large-scale partial differential equation (PDE) systems which 
need to be solved with, e.g., finite element methods. As such the total 
computational cost of the UQ tasks on these complex real-world systems can be prohibitive. 

A common remedy for reducing the computational cost is to construct computationally inexpensive 
approximation models 
and use them in the MC simulations. 
Loosely speaking, there are two main strategies to construct such approximation models.
The first one is to obtain a simpler (and thus less expensive) model by 
making certain simplification of the original model. 
Since the methods of this type typically require knowledge of the underlying model as well as the ability to modify it, 
they are usually referred to as being ``intrusive". 
A good example of the intrusive methods is the reduced-order modelling \cite{benner2017model}, which aims to reassemble the original model solution 
in a lower-dimensional space. 
In many real-world applications, however, it is often not practical to require the users to have the ability of analysing and modifying the underlying model. 
In this regard, an alternative class of approaches are to treat the underlying model as a black-box,
and construct the so-called emulator or surrogate only using the simulation data generated by the model. 
Since this type of methods only require the ability to simulate the underlying model as a black-box, 
without requiring any  knowledge of it, they are said to be ``non-intrusive''. 
Commonly used non-intrusive models include  radial basis functions, (collocation-based) polynomial chaos expansion, 
artificial neural networks, 
and Gaussian Process regression (GPR) ( see  \cite{ghanem2017handbook,smith2024uncertainty,sullivan2015introduction} for more discussions on the subject).
In this work we will focus on GPR as a surrogate model for the aforementioned UQ tasks. 
We will discuss both its advantages and limitations in this regard.

\section{An overview on the Gaussian Process regression}

Gaussian Process Regression (GPR) is a nonparametric and probabilistic method for regression. There are several ways to understand Gaussian Process (GP) regression models.
One approach is to view a Gaussian Process as defining a distribution over functions, with inference occurring directly in the space of functions. 
Another perspective is based on the joint Gaussian distribution and Bayesian inference. 
We here follow the joint Gaussian distribution viewpoint for  its simplicity and accessibility.

\subsection{Basics of the GPR model}
\label{sec:gp}
Given a real-valued function $f(\-x)$,  the GP or the Kriging method constructs a surrogate model of $f(\-x)$ in a nonparametric Bayesian regression framework~\cite{williams2006gaussian,gramacy2020surrogates}.
Specifically the target function $f(\-x)$ is cast as a Gaussian random process 
whose mean is $u_{\mathrm{pr}}(\-x)$ and covariance is specified by
a kernel function $k(\-x,\-x')$, namely,
\[ \COV[f(\-x),f(\-x')] = k(\-x,\-x'). \]
While in principle the prior mean can be any function, it is often convenient to use simple polynomials (e.g., constant, linear or quadratic functions). 
The kernel $k(\-x,\-x')$ is positive semidefinite and bounded. Now let us assume that $m$ evaluations of the function $f(\-x)$ are performed
 at parameter values $\-X^* := \left[\-x^*_1, \ldots \-x^*_m\right]$, yielding  function evaluations $\-y^* := \left[ {y}^*_1, \ldots {y}^*_m\right]$,
where \[{y}^*_i = f(\-x_i^*)\quad \mathrm{for} \quad i=1,\ldots,m.\]
Suppose that we want to predict the function value at location $\-x$, denoted as $y$.
The joint prior distribution of $(\-y^*,\,\-y)$ is, 
\begin{equation}  
\left[ \begin{array}{c}
         \-y^* \\
        y \end{array} \right] \sim \@N\left(\begin{array}{c}
         u_{\mathrm{pr}}(\-X^*) \\
        u_{\mathrm{pr}}(\-x) \end{array},
				\left[
				\begin{array}{ll}
         K(\-X^*,\-X^*) &K(\-X^*,\-x) \\
        K(\-x,\-X^*) &K(\-x,\-x) \end{array}\right]
				\right) , \label{e:jointdis}
\end{equation}
where we use the notation $K(\-A,\-B)$ to denote the matrix of the covariance evaluated at all pairs of points in set $\-A$ and in set $\-B$.
The posterior distribution of $\-y$ is also Gaussian: 
\begin{subequations}
\label{e:gp}
\begin{equation}
  y ~|~\-x,\-X^*, \-y^* \sim\mathcal{N}(u(\-x), \sigma^2(\-x)), \label{e:post}
\end{equation}
where the posterior mean is 
\begin{equation}
u(\-x)=u_{\mathrm{pr}}(\-x)+K(\-x,\-X^*)K(\-X^*,\-X^*)^{-1}(\-y^*-u_{\mathrm{pr}}(\-X^*)),\label{e:pmean}
\end{equation}
and the posterior variance is 
\begin{equation}
\label{eq:postcovsimple}
\sigma^2(\-x) = K(\-x,\-x)-K(\-x,\-X^*)K(\-X^*,\-X^*)^{-1}K(\-X^*,\-x). 
\end{equation}
\end{subequations}
  The posterior mean $u(\-x)$ in Eq.~\eqref{e:pmean} is usually used as the surrogate model 
  and the variance $\sigma^2(\-x)$ is used to provide uncertainty information in the value predicted by the surrogate model.  
  
  \textit{Note: Throughout this paper we reserve $u(x)$ and $\sigma^2(\-x)$ as the notation for the  mean and variance of a GP model.
Their precise meaning is specified by their superscripts and subscripts.}

In practice, the observed function values are often noisy. It is typically assumed that the observation noise is additive, expressed as \[ y^* = f(\mathbf{x}^*) + \eta ,\] 
where $\eta$ is assumed to follow a zero-mean Gaussian distribution with variance $\nu_{\text{obs}}^2$.
In this case, the prior on the noisy observations becomes a bit tricky: 
suppose that we have two observations $(x^*_i,y_i^*)$ and $(x_{j}^*,y_{j}^*)$, the prior covariance of $y^*_i$ and $y^*_j$ is
$$
\operatorname{COV}\left(y_i^*, y^*_j\right)=k\left(\mathbf{x}^*_i, \mathbf{x}^*_j\right)+\nu_{\text{obs}}^2 \delta_{i,j},
$$
where $\delta_{i,j}$ is a Kronecker delta which is one if $i=j$ and zero otherwise. 
It should be noted here that  $i$ and $j$ are the indices of observations: namely, $i=j$ implies 
that we consider the same data point,
and for  $i\neq j$, $x^*_i$ and $x^*_j$ are two independent observations. 
Introducing the noise term in Eq. \eqref{e:jointdis} we can write the joint distribution of the observed target values and the function values at the test locations under the prior
$$
\left[\begin{array}{l}
\mathbf{y}^* \\
{y}
\end{array}\right] \sim \mathcal{N}\left(\begin{array}{c}
         u_{\mathrm{pr}}(\-X^*) \\
        u_{\mathrm{pr}}(\-x) \end{array},
\left[\begin{array}{cc}
K(X^*, X^*)+\nu_{\text{obs}}^2 I & K\left(X^*, \-x\right) \\
K\left(\-x, X^*\right) & K\left(\-x, \-x\right)
\end{array}\right]\right).
$$
Deriving the conditional distribution corresponding to Eq.\eqref{e:gp}, we arrive at the predictive distribution for the noisy GP model: 
\begin{subequations}
\label{e:ngp}
\begin{align}
  y ~&|~\-D,\-X^*, \-y^* \sim\mathcal{N}(u(\-x), \sigma^2(\-x)), \\
 u(\-x) &= u_{\mathrm{pr}}(\-x)+K\left(\-x, \-X^*\right)\left[K(\-X^*, \-X^*)+\nu_{\text{obs}}^2 I\right]^{-1} \mathbf{y}^*,\\
\sigma^2(\-x) & =K\left(\-x, \-x\right)-K\left(\-x, \-X^*\right)\left[K(\-X^*, \-X^*)+\nu_{\text{obs}}^2 I\right]^{-1} K\left(\-X^*, \-x\right).
\end{align}
\end{subequations}

Finally it is worth noting that, the noise-free GPR can be considered as an interpolation method in the sense that it exactly fits all the data points;
in contrast, GPR with observation noise allows for fitting errors in the data points.
For UQ problems, the data points are typically generated via simulating the underlying model and thus are usually free of observation noise.
However, in practice, a small observation noise is often assumed to improve the stability of the GP computation, even when the data points are actually noise-free.

\subsection{Covariance kernels}
The performance of GPR depends critically on the choices of the covariance kernel functions
that measure the similarity between two data points. 
In this section, we will discuss several commonly used kernel functions, which can be classified into stationary and nonstationary types.
A stationary kernel can be expressed as a function of the difference between two points, i.e., $\boldsymbol{\tau}=\mathbf{x}-\mathbf{x}^{\prime}$,
while a nonstationary one depends on the actual values of $\-x$ and $\-x'$ rather than just their differences. 

Popular stationary kernels include:
\begin{itemize}
\item Squared Exponential (Gaussian) Kernel:
\begin{equation}
\label{eq:kernelSE}
k_{\text{SE}}(\-x, \-x') =  \sigma_0^2\exp\left(-\frac{\|\boldsymbol{\tau}\|^2}{2\ell^2}\right).
\end{equation}
Kernel parameters: 
\begin{itemize}
\item $\sigma_0^2$ controls the prior variance, 
\item parameter $\ell$ defining the characteristic length-scale which controls the rate of decay of similarity between points as their distance increases.
\end{itemize}
The Squared Exponential (SE) kernel is infinitely differentiable, which means that the GP with this covariance function has mean square derivatives of all orders, and is thus often used 
when the underlying function is (believed to be) very smooth.

\item Matérn Kernel:
\begin{equation}
k_{\text{Matérn}}(\-x, \-x') =\sigma_0^2\frac{2^{1-\kappa}}{\Gamma(\kappa)}\left(\frac{\sqrt{2\kappa}}{\ell} \|\boldsymbol{\tau}\|\right)^\kappa K_\kappa\left(\frac{\sqrt{2\kappa}}{\ell} \|\boldsymbol{\tau}\|\right).
\label{e:matern}
\end{equation}
where $\Gamma$ is the gamma function and $K_\kappa$ is the modified Bessel function of the second kind.  
Kernel parameters:
\begin{itemize}
\item $\sigma_0^2$ controls the prior variance, 
\item $\ell$ is the characteristic length-scale parameter, 
\item $\kappa>0$ controls the smoothness of the function, 
and is positive.   
\end{itemize}
The Matérn Kernel is often used in problems where the SE kernel is considered to be too smooth. 
In practice one often sets $\kappa=p+1 / 2$ with $p$ being a non-negative integer,
 as the kernel function can be simplified in this case. 
More specifically  $\kappa=3 / 2$ and $\kappa=5 / 2$ are commonly used values in GPR, yielding the kernel functions 
$$
\begin{aligned}
& k_{\kappa=3 / 2}(\-x,\-x')=\sigma_0^2\left(1+\frac{\sqrt{3} \|\boldsymbol{\tau}\|}{\ell}\right) \exp \left(-\frac{\sqrt{3} \|\boldsymbol{\tau}\|}{\ell}\right), \\
& k_{\kappa=5 / 2}(\-x,\-x')=\sigma_0^2\left(1+\frac{\sqrt{5} \|\boldsymbol{\tau}\|}{\ell}+\frac{5 \|\boldsymbol{\tau}\|^2}{3 \ell^2}\right) \exp \left(-\frac{\sqrt{5} \|\boldsymbol{\tau}\|}{\ell}\right). 
\end{aligned}
$$
Finally we note that, for $\kappa=1/2$, the Matérn kernel reduces to the exponential kernel,
\[k_{\text{EXP}}(\-x,\-x') =\sigma_0^2 \exp\left(-\frac{\|\boldsymbol{\tau}\|}l\right), \]
which is typically used to model ``rough''  functions, 
while  it converges to the SE kernel for $\kappa \rightarrow \infty$. 

\item Rational Quadratic (RQ) Kernel:
\[
k_{\text{RQ}}(\-x, \-x') = \sigma_0^2\left(1 + \frac{\|\boldsymbol{\tau}\|^2}{2\alpha\ell^2}\right)^{-\alpha}.
\]
Kernel parameters: 
\begin{itemize}
 \item $\sigma_0^2$ controls the prior variance, 
\item $\ell$ is the characteristic length-scale parameter, 
\item   $\alpha>0$ controls the smoothness of the function. 
\end{itemize}
The RQ kernel can be seen as a scale mixture (an infinite sum) of SE kernels with different characteristic length-scales. The limit of the RQ kernel for $\alpha \rightarrow \infty$ is the SE kernel with characteristic length-scale $\ell$.
\end{itemize}
A stationary kernel is said to be isotropic if it depends only on the distance $\|\boldsymbol{\tau}\|$. 
The aforementioned SE,  Matérn and RQ Kernels are all isotropic. 
These  kernels can be made anisotropic to handle data where different input dimensions may have different 
characteristic length-scales. For example, for two $d$-dimensional data points $\-x$ and $\-x'$, the anisotropic SE kernel is 
\[
k(\mathbf{x}, \mathbf{x'}) =  \sigma^2_{0}\exp \left( -\sum_{i=1}^d \frac{(x_i - x_i')^2}{2 l_i^2} \right),
\]
where $l_i$ is the characteristic length-scale for the $i$-dimension.
The anisotropic versions of other kernels can be obtained in a similar manner.

On the other hand, the nonstationary kernels can be used to model functions that exhibit varying smoothness properties at different locations in the input space. Some often-used nonstationary kernels include:
 \begin{itemize}
\item Linear Kernel:
\[
k_{\text{lin}}(\-x, \-x') = \-x^\top\-x'.
\]
GPR with the linear kernel is equivalent to the linear regression. 
\item Polynomial Kernel:
\[
k_{\text{poly}}(\-x, \-x') = (\mathbf{x}^\top \mathbf{x'} + c)^p,
\]
where $c\leq 0$  controls the trade-off between higher-order and lower-order terms, and $p$ is the polynomial order.
It is easy to see that when $d=1$ and $c=0$, $k_{\text{poly}}(\-x, \-x')$ reduces to the linear kernel. 
\end{itemize}

Model selection methods for choosing covariance kernels in GPR typically involve techniques like cross-validation (CV) and Bayesian model selection. These methods evaluate the performance of different kernels to identify the one that best captures the underlying structure of the data while balancing model complexity and predictive accuracy.
CV for model selection involves splitting the data into training and validation sets to evaluate the performance of different models, ultimately selecting the model that performs best on the validation data.
Bayesian model selection involves comparing models based on their posterior probabilities given the data, typically using Bayes factors or marginal likelihoods to quantify the evidence supporting each model.

\subsection{Hyper-parameter estimation}
As has been discussed in the previous section, most kernel functions depend on a number of hyperparameters 
 which often can not be predetermined. 
 In this regard,  methods that can determine these hyperparameter for a given data are important for GPR implementation. 
Here we introduce two common methods for estimating hyperparameters: maximum likelihood estimation (MLE) and cross validation (CV). 

We start with the MLE method. 
Namely,  let $\theta\in\Theta$ be the vector that collects all the hyperparameter in the kernel function, we determine $\theta$ by
\[\max_{\theta\in\Theta} \pi(\-y^*|\-X^*,\theta)\]

Next we let $l(\theta) = -\log \pi(\-y^*|\-X^*,\theta)$ and by neglecting some constant we obtain 
\begin{equation}
\min_{\theta\in\Theta} l(\theta)
:=\frac{1}{2} \mathbf{y^*}^{\top} \Sigma_y^{-1} \mathbf{y}^*+\frac{1}{2} \log \left|\Sigma_y\right|, \label{e:logll} \end{equation}
where $\Sigma_y=K(\-X^*, \-X^*)+\nu_{\text{obs}}^2 I$ is the covariance matrix for the noisy targets $\mathbf{y}^*$.
The first term in the right-hand-side of Eq.~\eqref{e:logll} 
is the data fidelity term measuring how $\Sigma_y$ fits the data,
and the second term  is the complexity penalty.
Standard gradient descent methods can be used to solve Eq.~\eqref{e:logll}, 
where we need to evaluate the gradient of $l(\theta)$. 
In this regard, the gradient of $l(\theta)$ can be derived analytically: 
\begin{equation}
\label{eq:diff_derivtive_MLL}
\frac{\partial}{\partial \theta_j} l(\theta)  =\frac{1}{2} \mathbf{y^*}^{\top} \Sigma^{-1} \frac{\partial \Sigma}{\partial \theta_j} \Sigma^{-1} \mathbf{y^*}-\frac{1}{2} \operatorname{tr}\left(\Sigma^{-1} \frac{\partial \Sigma}{\partial \theta_j}\right),
\end{equation}
where $\Sigma = K(\-X^*,\-X^*)$. 

CV is also widely used to estimate the hyperparameters of GP models. We here present the leave-one-out (LOO) CV method, while noting that the procedure can be readily extended to the more general $k$-fold CV. 
For $m=1,...,M$, let $\-X^*_{-m}$ and $\-y^*_{-m}$ be respectively $\-X^*$ with $\-x^*_m$ removed and $\-y^*$ with $y^*_m$ removed,
and  we construct the GP model with each $\-X^*_{-m}$ and $\-y^*_{-m}$:
$$\pi(y|\-x,\-X^*_{-m},\-y^*_{-m}) = \mathcal{N}(u_m(\-x),\sigma^2_m(\-x)),$$
where $u_m$ and $\sigma^2_m$ are the posterior mean and variance of the obtained GP model. 
Next we use the log predictive probability  as the  validation loss function:
\begin{equation}\label{e:lloo}
\begin{aligned}
\text{LLOO}(\theta) = -\sum_{m=1}^{M} \log \pi(y^*_m|\-x_m^*,\-X^*_{-m},\-y^*_{-m},\theta)\\
 = \frac{1}{2} \sum_{m=1}^M[\log \sigma^2_m +\frac{(y_i - u_m(\-x^*_m))^2}{\sigma^2_m}],
\end{aligned}
\end{equation}
where $\theta$ is also included as an argument variable of the distribution function to indicate the dependence of the GP model on it. 
Also note that a constant is neglected in the second equality of Eq.~\eqref{e:lloo}. 
The hyperparameter $\theta$ is then determined by minimizing the LLOO function. 
This can be done using gradient-based optimization techniques, and to this end its gradient can be computed as
\begin{equation}
\frac{\partial \text{LLOO}}{\partial \theta_j} = \sum_{n=1}^{N} 
\left( 
\alpha_n [Z_j \alpha]_n - \frac{1}{2} \left(1 + \frac{\alpha_n^2}{[\Sigma^{-1}]_{nn}}\right) [Z_j \Sigma^{-1}]_{nn} 
\right)
/[\Sigma^{-1}]_{nn},
\end{equation}
where \(\mathbf{\alpha} = \Sigma^{-1} \-y^*\) and \(Z_j = \Sigma^{-1} \frac{\partial \Sigma}{\partial \theta_j}\).

\subsection{Design of experiments in GPR} 
In many UQ related applications, the data points used for constructing GP models are often not available in advance;
 instead, users need to generate data by conducting simulations with the underlying model.
On the other hand, often the underlying models are highly expensive and one can only afford to generate 
a limited number of data points for constructing the GP model. 
As such determining the locations to query the underlying model  is a critical step in constructing GP surrogate for such models.
Querying the underlying models is often referred to as conducting computer experiments,
and therefore choosing in input locations of the computer experiments is  a question of experimental design. 

The first class of design approaches  seek design points that cover the input space (that is usually bounded) uniformly and thoroughly, ensuring that all regions are adequately represented. Such approaches are broadly referred to as space-filling designs.
Common space-filling design methods include Latin hypercube sampling (LHS), quasi-Monte Carlo, and quadrature.
These methods can largely be used for any function approximate techniques, and are not specifically tied to GPR. 

Another class of design methods sequentially determine the design points 
and in each step it selects the next design point 
that is expected to reduce the overall uncertainty in the GP model the most.
Since these methods make use of the uncertainty information, they are typically developed specifically for the GPR method. 
These methods are commonly known as  active learning methods, and in what follows we introduces two popular active learning methods. 
The first method uses the MacKay criterion (referred to as ALM) \cite{mackay1992information} which simply chooses the design point 
with the largest posterior variance. 
Another one uses the Cohn criterion (referred to as ALC) \cite{cohn1996active}, which measures the overall uncertainty in the GP model with the integrated posterior variance.
Next we provide the mathematical formulations for both. 


\textbf{ALM.} Suppose at iteration $N$, we have determined $m$ design points denoted as  $\-X^*_m$, and we seek the $(m+1)$-th. 
Suppose that the GP model is constructed with the dateset $\-X^*_m$,
its (posterior) variance is $\lambda_m(\-x)$,
and the active learning with MacKay criterion chooses the next point as, 
\[\-x^{\star}_{m+1}=\arg\max_{\-x \in D} \sigma^2_m(\-x)\]
where $D$ is the domain of $\-x$.

\textbf{ALC.}  Let $\-x$ be any point in $D$, and $\-X^*_{m+1}$ be the data set of adding $\-x$ to $\-X^*_m$. 
If we construct a GP model with data points on $\-X^*_{m+1}$, the resulting posterior variance is $\left(\sigma^{\-x}_{m+1}(\cdot)\right)^2$. 
It should be noted here that $\left(\sigma^{\-x}_{m+1}(\cdot)\right)^2$ can be computed without querying $f(\-x)$. 
The Cohn criterion finds the next point by maximizing the integrated posterior variance: 
\[\-x^{\star}_{m+1}=\arg\max_{\-x \in D}\int \left(\sigma^{\-x}_{m+1}(\-x')\right)^2\pi(\-x')d\-x'.\]
Compared with ALM, the ALC criterion considers the effect of each experiment on the entire domain and therefore yields better designs; ALC is more expensive, however, because it requires a new variance computation for each potential design. 

Active learning strategies have been an active research focus in the GP literature. In addition to ALM and ALC, many criteria have been proposed in more recent years. 
An example of these new criteria is the
     \textbf{Expected Improvement for Global Fit (EIGF)}~\cite{lam2008sequential}.     EIGF aims to improve the fit of the global GP surrogate by minimizing the prediction error in the overall GP:
    \[
   \-x^{\star}_{m+1}=\arg\max_{\-x\in D} \text{EIGF}(\mathbf{x}): = (u_m(\mathbf{x}) - f(\mathbf{x}^\dagger))^2 + \sigma_m^2(\mathbf{x}), \label{e:eigf}
    \]
    where  \( \mathbf{x}^{\dagger} \) is  the training point nearest to $\-x$.



\section{Applications in the UQ problems}

\subsection{GP for Uncertainty Propagation}
Recall that in the UP problem, the statistical quantity of interest is often calculated with the MC simulation given in Eq.~\eqref{e:upmc}. 
A straightforward application of GP in this setting is to first 
reconstruct the GP model of function $f(\-x)$, and use its posterior mean in Eq.~\eqref{e:upmc}, yielding:
\begin{equation}
E[f(\-x)]\approx\frac1n\sum_{n=1}^N u(\-x_n). \label{e:upgp}
\end{equation}
Eq.~\eqref{e:upgp} utilizes GP in a standard surrogate modeling formulation widely used for UQ problems~\cite{owen2017comparison}, 
in the sense that the same formulation applies to any non-intrusive surrogate models such as polynomial interpolation 
and radial basis functions. 
The formulation is straightforward and easy to implement, but it only uses the posterior mean of the GP model
and  does not take advantage of the uncertainty information obtained. 

Next we  consider an alternative method.  
We can see that the UP problem is essentially an integration problem -- we want to compute the integral 
\begin{equation}
I=\int f(\-x) \pi(\-x) d\-x. \label{e:fbar}
\end{equation}
Assume that we have $m$ data points $\-X^*$ and $\-y^*$, with which we construct a GP model $$\tilde{f}(\-x) \sim \mathcal{N}(u(\-x), \sigma^2(\-x)),$$ with $u(\-x)$ 
and $\sigma^2(\-x)$ given by Eq.~\eqref{e:pmean} and Eq.~\eqref{eq:postcovsimple} respectively. 
Now we replace $f(\-x)$ in Eq.~\eqref{e:fbar} by $\tilde{f}(\-x)$ yielding,
\begin{equation}
\tilde{I}=\int \tilde{f}(\-x) \pi(\-x) d\-x, \label{e:fbar2}
\end{equation}
where it should be clear that $\tilde{I}$ now is a random variable. 
It can be verified that the distribution of $\tilde{I}$ is also Gaussian, 
and its mean and variance can be derived analytically:
 \begin{equation}
 \begin{split}
 E[\tilde{I}] &= \int\int \tilde{f}(\-x)\pi(\-x)\mathrm{d} \-x\, \mu(d\tilde{f})  \\
                              &= \int u_{\mathrm{pr}}(\-x)\pi(\-x) d\-x +\-w K(\-X^*, \-X^*)^{-1} ( \mathbf{y}^* -u_{\mathrm{pr}}(\-X^*)),
 \end{split}
 \end{equation}
 \begin{equation}
 \begin{split}
 \text{VAR}[\tilde{I}] &= \int\left[\int f_{\text{GP}}(\-x)\pi(\-x)\mathrm{d} \-x - \int u(\-x')\pi(\-x')\mathrm{d} \-x'\right]^2 \mu(d\tilde{f})\\
                             & = \int \int k(\-x,\-x') \pi(\-x) \pi(\-x')d\-xd\-x'- \-w^\top K(\-X^*,\-X^*)^{-1}\-w  ,\label{3}
 \end{split}
 \end{equation}
 where $\-w=\left[\int K\left(\-X^*, \-x\right) \pi(\-x) d\-x\right]$ is a $m\times1$ vector
 with each component being $$w_i = \int k\left(\-x_j^*, \-x\right) \pi(\-x) d\-x \quad \mathrm{for}\quad j=1,...,m.$$ 
 The integrals in $w_i$ are called kernel mean embeddings. 

Next we consider the special case where $\pi(\cdot)$ is Gaussian (without loss of generality we can assume $\pi(\cdot)$ is a standard Gaussian), $k(\cdot,\cdot)$ is a SE kernel (see Eq.~\eqref{eq:kernelSE}) and $\mu$ is a polynomial function. 
We first note that in this case the integral $\int \mu(\-x)\pi(\-x) d\-x$ can be evaluated analytically and we will not discuss further details here. 
Each kernel mean embedding term $w_j$ also admits an analytical form: 
\[
w_j = \sigma_0^2 \left(\frac{\ell^2}{\ell^2 + 1}\right)^{d/2} \exp\left(-\frac{\|x^*_j\|^2}{2(\ell^2 + 1)}\right),
\]
where $d$ is the dimension of $\-x$. 
In this case the method is first proposed in \cite{o1991bayes} and termed as  Bayes-Hermite Quadrature or Bayesian Monte Carlo \cite{rasmussen2003bayesian}. 










It is worth mention that, when the underlying distribution is not Gaussian, an importance sampling approach can be employed. Specifically, the integral can be rewritten as:
\[  I = \int f(\-x)\pi(\-x)dx = \int f'(\-x)q(\-x)dx\]
where \( q(\-x) \) is a chosen Gaussian distribution and \( f'(\-x) = \frac{f(\-x)\pi(\-x)}{q(\-x)} \). 
This approach allows us, in principle, to evaluate the integral with
the Bayes-Hermite formulation, even when the original distribution \( \pi(\-x) \) is not Gaussian.
However, this treatment requires that the underlying distribution $\pi(\-x)$ (or even better, the product $f(\-x)\pi(\-x)$) can
be well approximated by a Gaussian distribution times a sufficiently smooth function. 
Otherwise, the resulting function $f'(\-x)$ will have rather complex structure and can not be easily approximated by a GP model. 
As such it is desirable to  deal with a general distribution $\pi(\-x)$ directly. 
One approach is to approximate the underlying distribution with a finite Gaussian mixture,
which can also yield analytical expressions \cite{kennedy1998bayesian}.

We also note that, within the present UQ context, if these integrals do not admit analytical expressions, they can 
be evaluated using standard numerical integration techniques such as MC. For example the MC estimator of $w_i$ is 
\[\hat{w}_j = \frac1N\sum_{n=1}^Nk(\-x^*_j,\-x_n),\]
where $\-x_n \sim \pi(\cdot)$ for $n=1,...,N$. 
 It is important to emphasize that these integrals can be accurately evaluated with reasonable computational cost, as they do not require evaluation of the original function \( f(\-x) \).
These methods, regardless of what the underlying distribution is, are commonly known as
Bayesian quadrature  (BQ) or Bayesian Monte Carlo. 
In practice, other techniques are often used in combination with BQ to further improve its performance,
e.g., control variate and multi-level methods.

A key challenge in BQ methods is experimental design—specifically, determining the points at which to query the underlying model \( f(x) \). It is important to note that our goal here is to enhance the accuracy of the integral estimate, which differs from the experimental design objective discussed in Section 2, where the focus is on constructing a globally accurate GP approximation of the underlying function.
Specifically, assuming that we have a dataset of $m$ points denoted as $\-X^*_m$,  we 
want to add a new point $\-x$, which should be chosen so that it maximally
reduces the variance $\text{Var}[\tilde{I}]$. 
Mathematically, the problem can be formulated as the following (after some elementary simplification): 
\[
    \max_{x\in D} U_{BQ}(\-x): = \-w_{m+1}^\top K(\-X_{m+1}^*,\-X_{m+1}^*)^{-1}\-w_{m+1} ,\]
    where
    \[\-w_{m+1}=\left[\int K\left(\-X_{m+1}^*, \-x'\right) \pi(\-x') d\-x'\right],
    \quad \text{and}\quad \-X_{m+1}^* = \-X^*_m\cup \{\-x\}.
\]
Interestingly, in the aforementioned experimental design formulation,  the active learning strategy is not useful, as the utility function 
is independent of the function values and so all the querying points can be determined in advance.
However, as is suggested in \cite{osborne2012active}, one can still employ the active learning strategy in this problem by including the hyperparameter optimization in each iteration.


 \subsection{GP for Risk Estimation}

Here we consider the RE problem described in Section \ref{sec:re}. 
Recall that the event of failure is defined as $f(\mathbf{x})<0$ and the failure probability is
$$
P=\mathbb{P}(f(\mathbf{x})<0)=\int_{\{ f(\mathbf{x})<0\}} \pi(\mathbf{x}) d \mathbf{x}=\int I_A(f(\mathbf{x})) \pi(\mathbf{x}) d \mathbf{x}. 
$$
where $A = (-\infty, 0)$.
The probability $P$ can be estimated via MC: 
\begin{equation}
    \hat{P} =  \frac1N \sum_{n=1}^N I_A(f(\-x_n)),
\end{equation}
where $x_1,...,x_N$ are drawn from $\pi(\-x)$. 

Just like the UP problems, when $f(\cdot)$ is expensive, we can construct the GP model and use it in the MC simulation. 
That is, we approximate the failure probability estimate as 
\[\hat{P} \approx \frac1N\sum_{n=1}^N I_A(u(\-x)),\]
where $u(\-x)$ is the posterior mean of the GP model of $f(\-x)$. 
While this class of methods is frequently employed in the literature, it presents a particular challenge in RE problems. Specifically, even 
when $u(\-x)$ is an accurate surrogate, the estimated probability may still exhibit significant error. This can be intuitively explained as follows: the estimation process can be viewed as a classification problem, where each sample is categorized as either belonging to the failure region or the safe region. However, there exists a region where the surrogate model misclassifies samples. If a significant number of samples fall into this misclassified region, the estimated failure probability will inevitably contain substantial error.
Therefore, when  constructing the GP model  in such problems, 
one must focus on accurately estimating the probability of failure rather than obtaining a globally accurate GP model.

A common strategy of constructing the GP model in RE problems is to use the active learning -- to sequentially 
identify the locations to query the true model, 
and this procedure is often combined with a sampling scheme. 
A common version of the combined algorithm is the Active-Kriging MC simulation (AK-MCS) method, which proceeds as the following:
\begin{enumerate}
    \item Generate a sufficiently large Monte Carlo population of \( N \) points:\\
    \( S = \{\-x_1, ..., \-x_N\} \).
    \item Randomly select \( N_1 \) points from \( S \), evaluated on the limit state function $f(\-x)$, 
    forming the initial training set.

     \item \textbf{(Active Learning Loop)}
    Repeat until a convergence criterion is satisfied:
    \begin{enumerate}
     \item Train a GP model using the current training set.
    \item Identify the next point \( x^{\star} \) in \( S \) to evaluate based on a prescribed utility function\footnote{The utility function  typically depends on the current 
    GP model.}  \( U(\-x) \):
    \[\-x^* = \arg\max_{\-x\in S} U(\-x), \]
    and add $(\-x^{\star}, f(\-x^{\star}))$ to the training set. 
    \end{enumerate}
        \item Use the Monte Carlo points in \( S \) and the current GP model to estimate the failure probability \( \hat{P} \).
    \end{enumerate}
Note that in the AK-MCS algorithm, often the underlying mathematical models
are complex and expensive and in the active learning loop,
one shall minimize the number of model evaluations needed. As such  
the choices of the utility function is crucial for the efficiency of the method.  
Below are some examples of the utility functions:
\begin{itemize}
\item The Expected Feasibility Function (EFF) \cite{bichon2008} is defined as:
\begin{equation}
\text{EFF}(x) = (\epsilon-|u(x)|) \int_{- \epsilon}^{ \epsilon}  \pi_{\tilde{f}(x)}(y) \, dy,
\end{equation}
where $\epsilon$ is a prescribed positive constant, $\tilde{f}(x)$ the current GP model, $u(x)$ is its mean and $\pi_{\tilde{f}(x)}$ is 
the probability density function (PDF) of the GP model at point $x$.
Intuitively EFF measures the expectation of a point $\-x$ is in the vicinity (characterized by parameter $\epsilon$) of the limit state (i.e., $f(\-x)=0$). 
 Since $\pi_{\tilde{f}(x)}(\cdot)$ is Gaussian whose mean is $u(\-x)$ and variance is $\sigma^2(\-x)$, EFF can be explicitly computed as 
\begin{multline}
\text{EFF}(\mathbf{x}) = u(\mathbf{x}) \left[ 2 \Phi\left(-\frac{ u(\mathbf{x})}{\sigma (\mathbf{x})}\right) 
- \Phi\left(\frac{ - \epsilon - u(\mathbf{x})}{\sigma (\mathbf{x})}\right)
- \Phi\left(\frac{ \epsilon - u(\mathbf{x})}{\sigma (\mathbf{x})}\right) \right]\\
- \sigma (\mathbf{x}) \left[ 2 \phi\left(\frac{ - u(\mathbf{x})}{\sigma (\mathbf{x})}\right) 
- \phi\left(\frac{- \epsilon - u(\mathbf{x})}{\sigma (\mathbf{x})}\right)
- \phi\left(\frac{ \epsilon - u(\mathbf{x})}{\sigma (\mathbf{x})}\right) \right]\\
+ \left[ \Phi\left(\frac{ \epsilon - u(\mathbf{x})}{\sigma (\mathbf{x})}\right) 
- \Phi\left(\frac{ - \epsilon - u(\mathbf{x})}{\sigma (\mathbf{x})}\right) \right]
\end{multline}
where $\Phi(\cdot)$ and $\phi(\cdot)$ are  the standard normal CDF and PDF respectively. 


\item Learning function $U$ \cite{echard2011ak} is directly defined using the GP mean and variance:
\[U(\-x)=\frac{|u(\-x)|}{\sigma(\-x)},\]
where $u(\-x)$ and $\sigma(\-x)$ are the mean and standard deviation of the current GP model. 
This utility function is constructed based on the exploration-exploitation criteria: 
 it balances \emph{exploration} by focusing on regions with high predictive uncertainty indicated by the 
 standard deviation of the GP model and 
 \emph{exploitation} by targeting regions where the mean prediction is closer to the limit state $0$.

 \item Stepwise Uncertainty Reduction (SUR)~\cite{bect2012sequential} treats the GP-based  estimate of the 
 failure probability as a random variable where the randomness comes from the GP model. 
 SUR defines the utility function as the expected  uncertainty reduction in the estimation by adding a point.
While noting that several versions of SUR proposed in \cite{bect2012sequential}, we here discuss one of them as an example. 
 Specifically, let $\-x$ be any point in the state space, $\-X^*_{m}$ be the current training set, and $\-X^*_{m+1}$ be the data set of adding $\-x$ to $\-X^*_m$. 
Now we assume that a GP model is constructed with data points on $\-X^*_{m+1}$,
denoted as $\tilde{f}^{\-x}_{m+1}(\cdot)$. 
Next one can define $\tau^{\-x}_{m+1}(\-x')$ as the probability that the GP model $\tilde{f}^{\-x}_{m+1}(\cdot)$ misclassifies point $\-x'$,
and it should be clear that 
 
  \[
 \text{SUR}(x) = \int  \tau^{\-x}_{n+1}(\-x') \, \pi(\-x')d\-x' ,
\]
where $\pi(\cdot)$ is the distribution of $\-x$.

\item The experimental design approach \cite{wang2016gaussian} formulates the  problem as an inference of the limit state, i.e., $\{\-z: f(\-z)=0\}$. 
Given that the current training set is D and the distribution for the limit state is $\pi(\-z|D)$. 
Now suppose we add a new data point $(\-x,\-y)$, and using Bayesian inference we have, 
$$\pi(\-z|\-x,\-y,D)= \frac{\pi(\-y|\-x,\-z,D)\pi(\-z|D)}{\pi(\-y|D)}.$$
We seek the new point $x$ that maximizes the information gain measured 
by the Kullback-Leibler divergence (KLD) between the posterior distribution  $\pi(z|x,y,D)$ and the prior distribution $\pi(z|D)$. 
\begin{align}
U(x) &= \int  D_{KL}(\pi(\-z|\-x,\-y,D), \pi(\-z|D))\pi(\-y|D) d\-y \notag \\
&= \int  \log \frac{\pi(\-z|\-x,\-y,D)}{\pi(\-z|D)} \pi(\-z|\-x,\-y,D)d\-z\pi(\-y|D) d\-y,
\end{align}
which can be simplified using the underlying GP model \cite{wang2016gaussian}. 
\end{itemize}
Note that, among the four example criteria provided, the first two can be computed analytically, whereas the last two require numerical integration.

When the sought probability is rare, the GP model is often incorporated with variance-reduction sampling schemes such as importance sampling, subset simulation, and Multi-canonical Monte Carlo (MC) to further improve computational efficiency. These combined methods can be categorized into two types: in the first type, the GP construction and the sampling scheme are largely implemented independently (see, e.g, \cite{echard2013combined}), and in the second type, the adaptation of the sampling scheme and the refinement of the GP model (i.e., the active learning) are performed interactively \cite{bect2017bayesian,wu2016surrogate}. Since detailed discussion of these methods requires a very detailed description of the sampling methods, which are beyond the scope of this tutorial, we therefore refer the readers to the aforementioned references for more comprehensive coverage of these techniques.

With the GP model, it is possible to evaluate the failure probability and also quantify the uncertainty in the estimation.
Now assume that a GP model has been constructed for the limit state function $f(\-x)$:
\[\tilde{f}(\-x) \sim N(u(\-x),\sigma^2(\-x)),\]
and we define the GP-based estimate of the failure probability as, 
 $$\tilde{P} = \int {I}_A(\tilde{f}(x)) \pi(\-x) d\-x.$$
It should be clear that $\tilde{P}$ is a random variable, reflecting the epistemic uncertainty due to the limited number of observations,
 and its mean and variance are \cite{dang2022structural}:
 \begin{subequations}\label{e:P_mean_var}
\begin{align} 
&\E[\tilde{P}]  
 =\int \Phi\left(-\frac{u_{f}({\-x})}{\sigma_{f}({\-x})}\right) \pi({\-x}) \mathrm{d} {\-x},\\
&\text{VAR}[\tilde{P}] 
= \int\int \Phi_2\left([\-0 \; \-0]; [u(\-x), u(\-x')], \Sigma_f(\-x, \-x')\right) \pi(\-x) \pi(\-x') \, d\-x \, d\-x' - (\E[\tilde{P}])^2,
\end{align}
\end{subequations}
where $\Phi(\cdot)$ is the standard normal CDF, $\Phi_{2}$ is the joint CDF of a bivariate normal distribution,
and $\Sigma_f(\-x,\-x')$ is the posterior covariance matrix of $\tilde{f}(\-x)$ and $\tilde{f}(\-x')$.
In practice Eqs.~\eqref{e:P_mean_var} need to be evaluated numerically.

\subsection{GP for Optimization under Uncertainty}
We now consider the OU problems, where the objective functions and/or the constraints are subject to uncertain factors. 
Following the SAA formulation in Section~\ref{sec:ou}, we assume that these functions are expensive black-box, and
 only noisy estimates of them are available. 
The GP model can be used to solve this type of optimization problems, resulting in the so-called Bayesian Optimization (BO) methods.
We refer to \cite{brochu2010tutorial,mockus2012bayesian,snoek2012practical} for more detailed discussion of BO. 

BO is a type of methods focused on solving the problem\footnote{Following the BO convention we here consider a maximization problem.}, 
\begin{equation}
\label{e:f_BO}
\max _{x \in A} f(x),
\end{equation}
where the objective and/or the constraint functions (represented by the feasible set $A$) are black-box, noisy and expensive to evaluate. 
As such typically no gradient information is available. 
Also BO is a search based global optimization technique in the sense that it seeks  a global rather than local optimum.
The method represents the objective function using a probabilistic surrogate model, typically a GP, which predicts the function's behavior across the search space. Just like many search based global optimization methods, BO iteratively selects the most promising points to evaluate based on a chosen criterion.

\begin{algorithm}[H]
    \caption{Basic  Bayesian Optimization algorithm}
    \label{alg:BO}
    \begin{algorithmic}[1]
    \STATE{\textbf{Inputs}:  a Gaussian process prior on $f$ and an acquisition function
     $a(x,\tilde{f})$ that depends on $x$ and GP model $\tilde{f}$.}
     \STATE{\textbf{Output}: the point evaluated with the largest $f(x)$.}
     \STATE Evaluate $f(\cdot)$ at $m_0$ points according to an initial space-filling experimental design,
     and denote the data set as $D_{m_0}=\{(x_i, y_i)\}_{i=1}^{m_0}$. Set $m=m_0$
\WHILE {$m\le M$}	
     \STATE Compute the GP model using data set $D_m$, denoted by $\tilde{f}_{m}$;
     \STATE Let 
      $x_{m+1} = \arg\max_{x\in A} a(\-x,\tilde{f}_{m})$;
     \STATE Evaluate $y_{m+1}=f\left(x_{m+1}\right)$ and let $D_{m+1}=D_m\cup\{(x_{m+1},y_{m+1})\}$
     \STATE $m=m+1$;
\ENDWHILE
   \end{algorithmic}
\end{algorithm}

We start with the case where the feasible set is  a simple set (e.g. a box constraint), in which it is easy to assess membership. A standard BO algorithm for this case proceeds as Alg.~\ref{alg:BO}. 
A key component of the BO algorithm is the acquisition function, which guides the selection of the next point to evaluate. Most of the acquisition functions
are constructed based on different strategies to balance the trade-off betwen exploration and exploitation:
loosely speaking exploitation focuses on refining solutions near known promising areas, while exploration seeks to gather more information about less-sampled regions to potentially discover better solutions.
Below are some common acquisition functions. 

\begin{itemize}
\item The Expected Improvement (EI) function aims to maximize the expected improvement over the current best observed value \( y_{\text{best}} \):
\[
\text{EI}(x) = \mathbb{E} \left[ \max(0, \tilde{f}(x) - y_{\text{best}}) \right]
\]
Using the expression of the GP model, this can be expressed as:
\[
\text{EI}(x) = (u(\-x) - y_{\text{best}}) \Phi(Z) + {\sigma(x)} \phi(Z)
\]
where \( Z = \frac{u(x) - y_{\text{best}}}{{\sigma(x)}} \),  \( \Phi(Z) \) and \( \phi(Z) \) 
are respectively the CDF and PDF of the standard normal distribution.

\item The Probability of Improvement (PI) function focuses on the likelihood of improving upon the current best value \( y_{\text{best}} \):
\[
\text{PI}(x) =P[\tilde{f}(x)\geq f^*+\xi)= \Phi \left( \frac{u(x) - y_{\text{best}} - \xi}{\sigma(x)} \right)
\]
where \( \xi \) is a parameter that adjusts the balance between exploration and exploitation.

\item The Upper Confidence Bound (UCB) function selects points by maximizing a confidence bound, 
\[
\text{UCB}(\-x) = u(\-x) + \xi \sigma(\-x)
\]
where \( \xi \) controls the balance between exploration and exploitation.
\end{itemize}





Next we discuss how to apply BO to  problems with rather complex constraints. First for complex equality constraints, they can typically be incorporated  into the objective function using standard  methods such as the penalty method.
As such we here mainly focus on handling inequality constraints. 
A main idea is to model each inequality constraint as a GP and include its information in the acquisition function as well. 
Most such methods use the  Probability of Feasibility (PoF) \cite{gardner2014bayesian} defined as the probability that a candidate solution $\-x$ satisfies the constraint.
Namely, suppose that the feasible set is defined as $A=\{ g(\-x)\leq 0\}$ where $g(\-x)$ is also a noisy and expensive black-box function,
and in this case we also construct a GP model for $g(\-x)$ in each BO iteration. 
Assuming that the GP model for $g$ is $\tilde{g} \sim N(u_g(x),\sigma^2_g(x))$, we can compute PoF as,  
\[ P_g(\-x) = P[ \tilde{g}(\-x)\leq 0]=
\Phi(-u_g(\-x)/{\sigma_g(\-x)}).\]
Now suppose that $a(\-x,\tilde{f})$ is the acquisition function for the objective function, and we can
\begin{itemize}
\item construct a new acquisition function for the constrained problem:
\[ a_{\text{con}}(\-x) = a(\-x,\tilde{f}) P_g(\-x).\]
\item optimize the acquisition function with a constraint on $P_g(x)$:
\[ \max_{x\in A} a(x,\tilde{f}) \quad\text{s.t.} \quad P_g(x)\geq \epsilon\]
for a prescribed threshold $\epsilon>0$.
\end{itemize}
Another popular method is to convert the problem into 
an unconstrained problem using the augmented Lagrangian formulation,
and we refer to \cite{gramacy2016modeling} for further details on this type of methods. 

Another interesting topic in BO research is the use of gradient information. In certain problems, the gradient of the objective function, or an estimate of it, is available. The challenge lies in determining how to effectively leverage this gradient information to enhance the performance of BO.
Methods in this category include leveraging gradient information to improve the GP construction, design new acquisition function, and conduct local search \cite{gao2020bayesian}.

\subsection{GP for Parameter Estimation}
As is discussed in Section \ref{sec:pe}, within the UQ context, 
we usually solve the PE problem in a Bayesian framework. 
Namely, given unknown parameter $\-x$ and observation $\-y$, we compute the posterior distribution $\pi(\-x|\-y)$ by Eq.~\eqref{e:poster}.
It has also been explained that usually one samples the posterior distribution using MCMC or other sampling methods, and in these methods, the likelihood function $\pi(\-y|\-x)$ must be queried repeatedly.


In many applications the likelihood function $\pi(\-y|\-x)$ 
involves a complex underlying mathematical model and is expensive to evaluate. 
Since the sampling scheme requires to query 
the likelihood function for many times, the total computational cost can be prohibitive. 
This is where GP enters the equation -- it is used as surrogate models for fast evaluation of the likelihood function. 
 A natural way of doing this is to model the log likelihood $\log \pi(\-y|\-x)$ as a GP.
This choice, however, does not include the prior distribution, which make
overlook important structure information encoded in the prior distribution. 
Therefore in practice, it is common to model the log joint distribution $\log \pi(\-x,\-y)$ a GP.
In what follows we define $$l(\-x) = \log \pi(\-x,\-y),$$ when observations $\-y$ are given. 
Note that the posterior distribution $\pi(\-x|\-y) \propto \exp( l(\-x))$, 
and if we have constructed a sufficiently accurate GP model $\tilde{l}(\-x)$ for $l(\-x)$,
we can approximate the unnormalized posterior as $\tilde{\pi}(\-x|\-y) \propto \exp(\tilde{l}(\-x))$. 
Finally we can conduct MCMC to draw samples from $\tilde{\pi}(\-x|\-y)$ (recall that we do not need the normalization constant in MCMC). 

 \begin{algorithm}[H]
    \caption{Bayesian Active Posterior Estimation}
 \label{alg:posterior_est}
    \begin{algorithmic}[1]
    \STATE{\textbf{Inputs}:  a Gaussian process prior on $l$ and an acquisition function
     $a(\mathbf{x},\tilde{l})$ that depends on $\mathbf{x}$ and GP model $\tilde{l}$.}
     \STATE{\textbf{Output}: a GP model for the log joint distribution ${l}(\mathbf{x})$.}
     \STATE Evaluate $l(\cdot)$ at $m_0$ points according to an initial space-filling experimental design,
     and denote the data set as $D_{m_0}=\{(\mathbf{x}_i, l_i)\}_{i=1}^{m_0}$. Set $m=m_0$
\WHILE {$m\le M$}	
     \STATE Compute the GP model using data set $D_m$, denoted by $\tilde{l}_{D_m}$;
     \STATE Let 
      $\mathbf{x}_{m+1} = \arg\max_{x\in A} a(\mathbf{x},\tilde{l}_{D_m})$ where $A$ is the state space of $x$;
     \STATE Evaluate $l_{m+1}=l\left(\mathbf{x}_{m+1}\right)$ and let $D_{m+1}=D_m\cup\{(\mathbf{x}_{m+1},l_{m+1})\}$
     \STATE $m=m+1$;
\ENDWHILE
   \end{algorithmic}
\end{algorithm}
We now discuss how to construct the GP model for $l(\-x)$. 
In this problem, the main task is to construct a sufficiently accurate GP model
why trying to reduce the number of evaluations of the target function $ l(\-x)$. 
This can also be done in an active learning fashion, which is particularly similar to the BO algorithm. 
Namely assume that we have already queried the the target function $l(\-x)$  at $m$ points: 
$l_i=l(\-x_i)$ for $i=1,...m$. 
Let $D_{m}=\left\{(\-x_i, l_i)\right\}_{i=1}^{m}$ be the set of these $m$ input output pairs.
We will identify the next querying point by maximizing a utility (or acquisition) function that usually depends on the current GP model constructed with $D_m$. 
A basic algorithm for this procedure, termed as Bayesian Active Posterior Estimation (BAPE) in \cite{kandasamy2015bayesian}, is given in Alg.~\ref{alg:posterior_est}. 
One can see that this algorithm is very similar to BO,
where the main difference lies on the utility function. 
In BO utility or acquisition functions are designed to identify the optimal solutions efficiently, whereas in Bayesian inference, they are constructed to achieve the most accurate approximation of the posterior distribution.
We here introduce two computationally tractable utility function for BAPE. 
\begin{itemize}
\item Exponentiated Variance (EV). Recall that in a standard GP construction, we often use variance as a measure of uncertainty, and choose to query the point with the highest variance under the current GP model. In the present problem, however, what is interested is not the GP model itself; rather it is 
the exponential function of it, which approximate the joint distribution. Therefore we should consider the variance of the exponentiated GP.
Specifically we take the utility function to be
$$
u_{\mathrm{EV}}\left(\-x\right)=\mathrm{Var}[\exp ( \tilde{l}(\-x))]=   \exp \left( 2 u(\-x)+\sigma^2\left(\-x\right)\right)\left(\exp \left(\sigma^2\left(\-x\right)\right)-1\right),
$$
where $u(\-x)$ and $\sigma^2(\-x)$ are the mean and variance of GP model $\tilde{l}(\-x)$. 
\item Exponentiated Entropy (EE). Since the exponentiated GP does not follow a Gaussian distribution, the variance may not be a suitable quantity for measuring the uncertainty. 
To this end, a more generic measure of uncertainty is the entropy. Therefore it is natural to use the entropy of the exponentiated GP as the utility function:
$$
u_{\mathrm{EE}}\left(\-x\right)=\mathbb{H}[\exp ( \tilde{l}(\-x))]=   u(\-x) + \frac12\log(2\pi e \sigma^2(\-x)).
$$
\end{itemize}

The basic BAPE scheme can be further improved for better performance. 
Below we discuss the adaptive GP (AGP) method as an example. 
In AGP one first writes the joint distribution $l(x)$ as
$$
\exp(l(\mathbf{x})) =\exp (g(\mathbf{x})) q(\mathbf{x}),
$$
where $q(\mathbf{x})$ is a probability distribution that we are free to choose.
It is easy to see that
\begin{equation}
\label{eq:para_est_expensive}
g(\mathbf{x})=l(\mathbf{x})-\log (q(\mathbf{x})),
\end{equation}
and we will construct the GP model for $g(\mathbf{x})$.
 It should be noted that the distribution $q(\mathbf{x})$ plays an important role in the surrogate construction, as a good choice of $q(\mathbf{x})$ can significantly improve the accuracy of the GP  model. In particular, if we take $q(\mathbf{x})$ to be exactly the posterior $p(x|y)$, it follows immediately that $g(\mathbf{x})$ in Eq.~\eqref{eq:para_est_expensive} is a constant. This then gives us the intuition that if $q(\mathbf{x})$ is a good approximation to the posterior,  $g(\mathbf{x})$ is a mildly varying function that is easy to approximate. In other words, we can improve the performance of the GP surrogate by factoring out a good approximation of the posterior. Certainly this cannot be done in one step, as the posterior is not known in advance.
 We present an adaptive framework to construct a sequence of pairs $\left\{q_n(\mathbf{x}), \exp \left(\tilde{g}_n(\mathbf{x})\right)\right\}_{n=0}^N$, the product of which evolves to a good approximation of the joint distribution $\exp(l(\mathbf{x}))$.  \
 Roughly speaking, in the $n$th iteration, given the current GP model is $\tilde{g}_n(\mathbf{x})$, 
 and the approximate posterior is $q_n(x)$, the AGP algorithm performs the following: 
 \begin{enumerate}
\item use MCMC to draw a set of samples from the  approximate posterior $\exp(\tilde{g}_n(\mathbf{x}))q_n(\mathbf{x})$;
\item select one or multiple querying points based on a chosen utility function; evaluate 
the $l(\mathbf{x})$ at the selected point(s) and update the training set;
\item conduct a density estimation of the samples and obtain a new approximate posterior $q_{n+1}(\mathbf{x})$;
\item construct the GP model for 
$$g_{n+1}(\mathbf{x}) = l(\mathbf{x})-\log q_{n+1}(\mathbf{x}),$$
with the updated training set.
 \end{enumerate}

\subsection{GP for sensitivity analysis}
Recall that sensitivity analysis focuses on quantifying the relative importance of input factors in contributing to the variability of the model outputs and identifying parameters that are critical for system performance.
As has been discussed in Section \ref{sec:sa}, 
 the sensitivity indices are computed using the MC sampling. 
The motivation of using the GP model is same as before: computing these indices require a large number simulations of the underlying model $f(x)$ which are often computationally intensive, and replacing it with an accurate GP surrogate can significantly reduce the total computational cost. 

The basic scheme  to use GP in global SA (GSA) problems is straightforward: one first constructs a sufficiently accurate GP model for function $f(x)$ using a training set generated by a space filling method, and then uses the GP model (more precisely its mean) in the MC simulations. That is, Eq.~\eqref{e:eval_f} becomes, 
\begin{equation}
Y_A = u(A), \quad Y_B = u(B), \quad Y_{A_B^{(i)}} = u(A_B^{(i)}), \label{e: eval_gp}
\end{equation}
where $u(\cdot)$ is the mean of the GP model. The data obtained are then used in Eqs.~\eqref{eq:mcsa} to compute the SA indices. 
An advantage of the GP method is that we can also compute the confidence interval of the estimates, quantifying the uncertainty in them. 
In the basic version of the GP based GSA, the GP model is pre-built using some space-filling techniques. 
Similar to other UQ problems, a natural improvement of the basic version is to adaptively refine the GP model using an active learning strategy.
The refining procedure is also similar to those used in other UQ problems and namely in each iteration, we do the following, 
\begin{enumerate}
\item select one or more querying points  based on a utility function;
\item evaluate the function $f(\mathbf{x})$ at the selected point(s) and add them in the training set;
\item update the GP model with the new training set.
\end{enumerate}
The key in the method is to design a utility function that identifies points most effective in improving the estimation of the SA indices.
Here we present such a utility function developed in \cite{chauhan2024active}: 
    {Minimize Uncertainty in Sobol Index Convergence (MUSIC)}, 
which is specifically designed for the SA indices.  
Mathematically the MUSIC utility is
    \[
    \text{MUSIC}(\mathbf{x}) = \mathbf{D}^T(\mathbf{x}) \mathbf{M}(\mathbf{x}),
    \]
    where \( \mathbf{D}(\mathbf{x}) \) is a distance-based prefactor for balancing contributions across dimensions,
    and  \( \mathbf{M}_A(\mathbf{x}) \) is the vector representing the utility functions for all dimensions 
    (namely the $i$-th component of \( \mathbf{M}_A(\mathbf{x}) \) is the utility function for the $i$-th dimension of $\-x$). 
 The authors of \cite{chauhan2024active} give two possible choices of   \( \mathbf{D}(\mathbf{x}) \):
\[
\mathbf{D}_1(\mathbf{x}) = \mathbf{W} \cdot 
\begin{bmatrix}
|x_1 - x_1^{(j^*)}| \\
|x_2 - x_2^{(j^*)}| \\
\vdots \\
|x_d - x_d^{(j^*)}|
\end{bmatrix}
\quad \text{and} \quad 
\mathbf{D}_2(\mathbf{x}) = \mathbf{W} \cdot \| \mathbf{x} - \mathbf{x}^{(j^*)} \|_2^2 \mathbf{1}_d,
\]
where
 \( \mathbf{W} = [w_1, w_2, \dots, w_d]^T \) is a user-chosen weight vector with \( \sum_{i=1}^d w_i = 1 \),
 and \( x_i^{(j^*)} \) is the \( i \)-th component of the nearest training sample to \( \-x \).
 The utility  vector \( \mathbf{M}(\mathbf{x}) \) can be constructed by applying the EIGF utility function~\eqref{e:eigf}
 to each dimension. Namely, assuming that the mean and variance of the current GP model is $u(x)$ and $\sigma^2(x)$ respectively, 
 we define the function \( \mathbf{M}(\mathbf{x}) \) as
 \[
\mathbf{M}(\mathbf{x}) =
\begin{bmatrix}
\mathrm{EIGF}_{1}(x_1)] \\
\mathrm{EIGF}_{2}(x_2)] \\
\vdots \\
\mathrm{EIGF}_{n}(x_1)]
\end{bmatrix}
\]
 where 
    \begin{align*}
&    \mathrm{EIGF}_i(x_i) = (\mu_{i}(x_i) - \mu_{i}(x_i^{(j^*)}))^2 + \upsilon_{i}^2(x_i),\\
& \mu_i(x_i) = \E_{\pi(\-x_{-i}|x_i)}[u(\-x)|x_i], \\
&  \upsilon^2_i(x_i) = \mathrm{Var}_{\pi(\-x_{-i}|x_i)}[u(\-x)|x_i],
\end{align*}
and $\pi(\-x_{-i}|x_i)$ is  the conditional distribution of all input variables excluding $x_i$ (denoted as $\-x_{-i}$)
given a fixed value of $x_i$. 
In practice, $\mu_i(x_i)$ and  $\upsilon^2_i(x_i)$ are estimated with the MC method.

\section{Discussions}

As has been demonstrated, the GP methods have been widely used in  UQ problems. 
Here we briefly summarize the advantages and limitations of the UQ methods. 

\subsection{Advantages}

\paragraph{Flexibility – Nonparametric Nature} 
One of the primary strengths of GPs is their nonparametric formulation, allowing them to adapt to a wide variety of data structures without the need to predefine the complexity of the model. Unlike parametric models, which have a fixed number of parameters, GPs can flexibly grow their capacity as data increases, making them well-suited for modeling complex functions with minimal assumptions.
That said, the GP method is not unlimited in its flexibility. Its performance still depends on the model adopted and, most importantly, the choice of the kernel function. Selecting the most appropriate kernel function is crucial to achieving the best performance, as it encodes assumptions about the underlying structure of the data. 

\paragraph{UQ capacity – Probabilistic Formulation} 
GPs naturally provide uncertainty quantification, which is embedded in their probabilistic framework. Most importantly, they can accommodate uncertainty arising from insufficient data, providing a clearer representation of where the model lacks confidence. 
The ability to estimate uncertainties is invaluable for tasks where understanding the confidence of predictions is crucial, as it
not only enhances model interpretability but also enables better decision-making under uncertainty.
However, it is also important to note that the uncertainty information represented by the GP model may not always be accurate. The reliability of these estimates depends on factors such as the quality of the data, the choice of the model (e.g., the kernel function), and the validity of the underlying assumptions in the model.

\paragraph{Experimental Design and Active Learning} 
As is mentioned earlier, the GP model can characterize the uncertainty in the estimates corresponding to different inputs, which provides
useful information to effectively guide data acquisition,
and enables experimental design formulation for efficient construction of the surrogate model. 
By leveraging uncertainty estimates, GPs can identify regions in the input space where additional data is most valuable, leading to faster convergence with fewer samples. This makes them particularly effective in scenarios where data collection is expensive or time-consuming.

\subsection{Limitations}

\paragraph{Dimensionality} 
A significant limitation of GPs is their sensitivity to high-dimensional input spaces. As the dimensionality increases, the performance of GPs can degrade due to the curse of dimensionality.
In high-dimensional spaces, data points tend to become sparse, making it difficult for GPs to learn meaningful patterns from the data. Kernel functions, which are central to GPs, may struggle to adequately capture relevant structure or similarity between data points in such spaces, often resulting in poor model performance or overfitting.
Additionally high-dimensional input spaces may also exacerbate issues with hyperparameter optimization for the kernel, as the search space becomes more complex and harder to navigate.
These issues underscore the importance of dimensionality reduction techniques  when applying GPs to such problems \cite{tripathy2016gaussian,xiong2022clustered,chen2020anova}.

\paragraph{Scalability with Large Datasets} 
Models with high input dimensionality and complex structures often require a substantial amount of data to construct a reliable GP model. However, GPs face significant computational challenges when dealing with large datasets, as the training and inference involve matrix operations with cubic complexity. This makes scaling GPs to large datasets both computationally expensive and memory-intensive, making
them infeasible for large-scale and high complex models. 
There are certain techniques, notably sparse approximations and inducing point methods\cite{quinonero2005unifying,snelson2005sparse,titsias2009variational,hensman2013gaussian,wilson2015kernel}, that can mitigate this issue to some extent; however, these approaches rely on approximations and may compromise computational accuracy.
This issue requires further study, and the development of more efficient and accurate solutions for scaling GPs to large datasets is highly needed.

\subsection{Comparison with Neural Network Models}
Thanks to the rapid advancement of computational power, deep neural network (DNN) based machine learning models have become unprecedentedly popular for 
describing and chracterizing large-scale and complex systems \cite{willard2022integrating,karniadakis2021physics}. 
Some of such models have been directly applied to UQ problems \cite{tripathy2018deep}.
This raises a very natural question: in such a circumstance, is the GP method, with its major limitations just mentioned, still a useful surrogate model, or has it been (or will it be) largely superseded by the DNNs?
Our view on this question is that GPs and DNN are, in some sense, complementary rather than competing methodologies. The reason is two-fold.

First the two methods are suited for
dramatically different problem settings. 
DNNs excel in high-dimensional and large-scale data scenarios, benefiting from their scalability and ability to learn 
complex model structures.
On the other hand,  DNN methods 
are highly demanding in terms of computational resource and data, and thus may not be suitable for problems where these requirements can not be met. 
To this end, GPs offer important advantages in problems where
data is scarce, and computational efficiency is critical.  
First of all, the training of GP models are are typically efficient, especially 
when the mount of data is not large.
More importantly, in such problems, due to the data scarcity, the surrogate models are inevitably subject to uncertainty, and as such GPs' ability to quantify uncertainty is also highly valuable. 

Furthermore, GPs can be  integrated with DNN models in hybrid approaches. 
Specifically, the DNN model can be used to construct a comprehensive latent representation or surrogate model, without being constrained
to a specific model setting or data input.
Such a comprehensive model is developed during an offline stage, in which we assume that sufficient computational power is available, and computational time is less of a constraint. During the online stage, 
information on input data and model setting become available, 
and the pre-built DNN models may need to be refined or corrected based on such information. 
 At this stage, computational resources and data may be limited, 
 and responses need to be generated in real-time. 
 In this regard,
the GP model can be used as a ``last-mile'' tool  that connects the pre-trained DNN model with the real-time information based prediction/decision making process,
thanks to its computational efficiency and uncertainty quantification capacity.

\section*{Acknowledgement} This work is adapted from a short course that the authors taught at the Chinese Academy of Sciences (CAS) in the summer of 2024. We are grateful to CAS for their hospitality and generous support of the course.

{\small
\bibliographystyle{plain}
\bibliography{gp0,gp1}

\begin{thebibliography}{10}

\bibitem{bect2012sequential}
Julien Bect, David Ginsbourger, Ling Li, Victor Picheny, and Emmanuel Vazquez.
\newblock Sequential design of computer experiments for the estimation of a
  probability of failure.
\newblock {\em Statistics and Computing}, 22:773--793, 2012.

\bibitem{bect2017bayesian}
Julien Bect, Ling Li, and Emmanuel Vazquez.
\newblock Bayesian subset simulation.
\newblock {\em SIAM/ASA Journal on Uncertainty Quantification}, 5(1):762--786,
  2017.

\bibitem{benner2017model}
Peter Benner, Mario Ohlberger, Albert Cohen, and Karen Willcox.
\newblock {\em Model reduction and approximation: theory and algorithms}.
\newblock SIAM, 2017.

\bibitem{bichon2008}
Barron~J Bichon, Michael~S Eldred, Laura~Painton Swiler, Sandaran Mahadevan,
  and John~M McFarland.
\newblock Efficient global reliability analysis for nonlinear implicit
  performance functions.
\newblock {\em AIAA journal}, 46(10):2459--2468, 2008.

\bibitem{brochu2010tutorial}
Eric Brochu, Vlad~M Cora, and Nando De~Freitas.
\newblock A tutorial on bayesian optimization of expensive cost functions, with
  application to active user modeling and hierarchical reinforcement learning.
\newblock {\em arXiv preprint arXiv:1012.2599}, 2010.

\bibitem{chauhan2024active}
Mohit~S Chauhan, Mariel Ojeda-Tuz, Ryan~A Catarelli, Kurtis~R Gurley, Dimitrios
  Tsapetis, and Michael~D Shields.
\newblock On active learning for gaussian process-based global sensitivity
  analysis.
\newblock {\em Reliability Engineering \& System Safety}, 245:109945, 2024.

\bibitem{chen2020anova}
Chen Chen and Qifeng Liao.
\newblock Anova gaussian process modeling for high-dimensional stochastic
  computational models.
\newblock {\em Journal of Computational Physics}, 416:109519, 2020.

\bibitem{cohn1996active}
David~A Cohn, Zoubin Ghahramani, and Michael~I Jordan.
\newblock Active learning with statistical models.
\newblock {\em Journal of artificial intelligence research}, 4:129--145, 1996.

\bibitem{dang2022structural}
Chao Dang, Marcos~A Valdebenito, Matthias~GR Faes, Pengfei Wei, and Michael
  Beer.
\newblock Structural reliability analysis: A bayesian perspective.
\newblock {\em Structural Safety}, 99:102259, 2022.

\bibitem{echard2011ak}
Benjamin Echard, Nicolas Gayton, and Maurice Lemaire.
\newblock Ak-mcs: an active learning reliability method combining kriging and
  monte carlo simulation.
\newblock {\em Structural safety}, 33(2):145--154, 2011.

\bibitem{echard2013combined}
Benjamin Echard, Nicolas Gayton, Maurice Lemaire, and Nicolas Relun.
\newblock A combined importance sampling and kriging reliability method for
  small failure probabilities with time-demanding numerical models.
\newblock {\em Reliability Engineering \& System Safety}, 111:232--240, 2013.

\bibitem{gao2020bayesian}
Yuzhou Gao, Tengchao Yu, and Jinglai Li.
\newblock Bayesian optimization with local search.
\newblock In {\em Machine Learning, Optimization, and Data Science: 6th
  International Conference, LOD 2020, Siena, Italy, July 19--23, 2020, Revised
  Selected Papers, Part II 6}, pages 350--361. Springer, 2020.

\bibitem{gardner2014bayesian}
Jacob~R Gardner, Matt~J Kusner, Zhixiang~Eddie Xu, Kilian~Q Weinberger, and
  John~P Cunningham.
\newblock Bayesian optimization with inequality constraints.
\newblock In {\em ICML}, volume 2014, pages 937--945, 2014.

\bibitem{gelman2013bayesian}
Andrew Gelman, John~B. Carlin, Hal~S. Stern, David~B. Dunson, Aki Vehtari, and
  Donald~B. Rubin.
\newblock {\em Bayesian Data Analysis}.
\newblock CRC Press, Boca Raton, FL, 3rd edition, 2013.

\bibitem{ghanem2017handbook}
Roger Ghanem, David Higdon, Houman Owhadi, et~al.
\newblock {\em Handbook of uncertainty quantification}, volume~6.
\newblock Springer New York, 2017.

\bibitem{gramacy2020surrogates}
Robert~B Gramacy.
\newblock {\em Surrogates: Gaussian process modeling, design, and optimization
  for the applied sciences}.
\newblock Chapman and Hall/CRC, 2020.

\bibitem{gramacy2016modeling}
Robert~B Gramacy, Genetha~A Gray, S{\'e}bastien Le~Digabel, Herbert~KH Lee,
  Pritam Ranjan, Garth Wells, and Stefan~M Wild.
\newblock Modeling an augmented lagrangian for blackbox constrained
  optimization.
\newblock {\em Technometrics}, 58(1):1--11, 2016.

\bibitem{hensman2013gaussian}
James Hensman, Nicolo Fusi, and Neil~D Lawrence.
\newblock Gaussian processes for big data.
\newblock {\em arXiv preprint arXiv:1309.6835}, 2013.

\bibitem{kandasamy2015bayesian}
Kirthevasan Kandasamy, Jeff Schneider, and Barnab{\'a}s P{\'o}czos.
\newblock Bayesian active learning for posterior estimation.
\newblock In {\em Proceedings of the 24th International Conference on
  Artificial Intelligence}, pages 3605--3611, 2015.

\bibitem{kennedy1998bayesian}
Marc Kennedy.
\newblock Bayesian quadrature with non-normal approximating functions.
\newblock {\em Statistics and Computing}, 8:365--375, 1998.

\bibitem{kim2015guide}
Sujin Kim, Raghu Pasupathy, and Shane~G Henderson.
\newblock A guide to sample average approximation.
\newblock {\em Handbook of simulation optimization}, pages 207--243, 2015.

\bibitem{lam2008sequential}
Chen~Quin Lam.
\newblock {\em Sequential adaptive designs in computer experiments for response
  surface model fit}.
\newblock PhD thesis, The Ohio State University, 2008.

\bibitem{mackay1992information}
David~JC MacKay.
\newblock Information-based objective functions for active data selection.
\newblock {\em Neural computation}, 4(4):590--604, 1992.

\bibitem{mockus2012bayesian}
Jonas Mockus.
\newblock {\em Bayesian approach to global optimization: theory and
  applications}, volume~37.
\newblock Springer Science \& Business Media, 2012.

\bibitem{o1991bayes}
Anthony O'Hagan.
\newblock Bayes--hermite quadrature.
\newblock {\em Journal of statistical planning and inference}, 29(3):245--260,
  1991.

\bibitem{osborne2012active}
Michael Osborne, Roman Garnett, Zoubin Ghahramani, David~K Duvenaud, Stephen~J
  Roberts, and Carl Rasmussen.
\newblock Active learning of model evidence using bayesian quadrature.
\newblock {\em Advances in neural information processing systems}, 25, 2012.

\bibitem{owen2017comparison}
Nicola~E Owen, Peter Challenor, Prathyush~P Menon, and Samir Bennani.
\newblock Comparison of surrogate-based uncertainty quantification methods for
  computationally expensive simulators.
\newblock {\em SIAM/ASA Journal on Uncertainty Quantification}, 5(1):403--435,
  2017.

\bibitem{quinonero2005unifying}
Joaquin Quinonero-Candela and Carl~Edward Rasmussen.
\newblock A unifying view of sparse approximate gaussian process regression.
\newblock {\em The Journal of Machine Learning Research}, 6:1939--1959, 2005.

\bibitem{rasmussen2003bayesian}
Carl~Edward Rasmussen and Zoubin Ghahramani.
\newblock Bayesian monte carlo.
\newblock {\em Advances in neural information processing systems}, pages
  505--512, 2003.

\bibitem{robert2004monte}
Christian~P. Robert and George Casella.
\newblock {\em Monte Carlo Statistical Methods}.
\newblock Springer Texts in Statistics. Springer, New York, NY, 2nd edition,
  2004.

\bibitem{rubinstein2016simulation}
Reuven~Y Rubinstein and Dirk~P Kroese.
\newblock {\em Simulation and the Monte Carlo method}.
\newblock John Wiley \& Sons, 2016.

\bibitem{saltelli2008global}
A~Saltelli.
\newblock {\em Global Sensitivity Analysis: the Primer}.
\newblock John Wiley \& Sons, 2008.

\bibitem{smith2024uncertainty}
Ralph~C Smith.
\newblock {\em Uncertainty quantification: theory, implementation, and
  applications}.
\newblock SIAM, 2024.

\bibitem{snelson2005sparse}
Edward Snelson and Zoubin Ghahramani.
\newblock Sparse gaussian processes using pseudo-inputs.
\newblock {\em Advances in neural information processing systems}, 18, 2005.

\bibitem{snoek2012practical}
Jasper Snoek, Hugo Larochelle, and Ryan~P Adams.
\newblock Practical bayesian optimization of machine learning algorithms.
\newblock In {\em Advances in neural information processing systems}, pages
  2951--2959, 2012.

\bibitem{sullivan2015introduction}
Timothy~John Sullivan.
\newblock {\em Introduction to uncertainty quantification}, volume~63.
\newblock Springer, 2015.

\bibitem{titsias2009variational}
Michalis Titsias.
\newblock Variational learning of inducing variables in sparse gaussian
  processes.
\newblock In {\em Artificial intelligence and statistics}, pages 567--574.
  PMLR, 2009.

\bibitem{tripathy2016gaussian}
Rohit Tripathy, Ilias Bilionis, and Marcial Gonzalez.
\newblock Gaussian processes with built-in dimensionality reduction:
  Applications to high-dimensional uncertainty propagation.
\newblock {\em Journal of Computational Physics}, 321:191--223, 2016.

\bibitem{wang2016gaussian}
Hongqiao Wang, Guang Lin, and Jinglai Li.
\newblock Gaussian process surrogates for failure detection: A bayesian
  experimental design approach.
\newblock {\em Journal of Computational Physics}, 313:247--259, 2016.

\bibitem{williams2006gaussian}
Christopher~KI Williams and Carl~Edward Rasmussen.
\newblock {\em Gaussian processes for machine learning}.
\newblock MIT Press, 2006.

\bibitem{wilson2015kernel}
Andrew Wilson and Hannes Nickisch.
\newblock Kernel interpolation for scalable structured gaussian processes
  (kiss-gp).
\newblock In {\em International conference on machine learning}, pages
  1775--1784. PMLR, 2015.

\bibitem{wu2016surrogate}
Keyi Wu and Jinglai Li.
\newblock A surrogate accelerated multicanonical monte carlo method for
  uncertainty quantification.
\newblock {\em Journal of Computational Physics}, 321:1098--1109, 2016.

\bibitem{xiong2022clustered}
Junda Xiong, Xin Cai, and Jinglai Li.
\newblock Clustered active-subspace based local gaussian process emulator for
  high-dimensional and complex computer models.
\newblock {\em Journal of Computational Physics}, 450:110840, 2022.

\end{thebibliography}
}

\end{document}